\documentclass[preprint,superscriptaddress,amsmath,amssymb,natbib, caption, nofootinbib]{revtex4-1}
\usepackage{graphicx}

\begin{document}

\title{Design Overview of the DM Radio Pathfinder Experiment}

\author{$^1$Maximiliano Silva-Feaver, $^1$Saptarshi Chaudhuri, $^3$Hsaio-Mei Cho, $^2$Carl Dawson, $^1$Peter Graham, $^{1,3}$Kent Irwin, \\$^1$Stephen Kuenstner, $^3$Dale Li, $^1$Jeremy Mardon, $^5$Harvey Moseley, $^2$Richard Mule, $^1$Arran Phipps, $^4$Surjeet Rajendran, \\$^2$Zach Steffen, $^2$Betty Young\\
$^1$Stanford University, 
Stanford, California 94305\\ $^2$Santa Clara University, Santa Clara, California 95053\\ $^3$SLAC National Accelerator Laboratory, Menlo Park, California 94025\\ $^4$University of California Berkeley, Berkeley, California 94720\\$^5$NASA Goddard Space Flight Center, Greenbelt, Maryland 20771}

\begin{abstract}
We introduce the DM Radio, a dual search for axion and hidden photon dark matter using a tunable superconducting lumped-element resonator. We discuss the prototype DM Radio Pathfinder experiment, which will probe hidden photons in the 500 peV (100 kHz)-50 neV (10 MHz) mass range. We detail the design of the various components: the LC resonant detector, the resonant frequency tuning procedure, the differential SQUID readout circuit, the shielding, and the cryogenic mounting structure. We present the current status of the pathfinder experiment and illustrate its potential science reach in the context of the larger experimental program.
\end{abstract}

\maketitle

\section{Introduction}
Over the past three decades, there have been numerous experimental efforts aimed at the direct detection of dark matter, with the greatest focus on the theoretically favored Weakly Interacting Massive Particle (WIMP). So far, all experiments have returned null results despite having searched over a large range of mass and interaction cross-sections. Recently, there has been increased theoretical and experimental interest in searching for ultra-light dark matter candidates. \cite{snowmass}-\cite{CabreraTalk} Chief among these candidates are the axion and the hidden photon. The axion is a pseudoscalar, spin-0 particle. In addition to being a natural dark matter candidate (owing to its theoretically predicted small couplings to baryonic matter), it may also solve the strong CP problem, which addresses the apparent absence of CP violation in quantum chromodynamics\cite{PecceiQuinn}. The hidden photon is a vector, spin-1 particle that naturally appears in many extensions of the Standard Model. Hidden-photon dark matter would likely be generated by cosmic inflation\cite{Inflationary Fluctuations}.

Both the axion and hidden photon are predicted to have a small, but nonzero, coupling to photons. DM Radio is an experiment to search for hidden photons and axions using a tunable superconducting lumped-element resonator. It will instrument $\sim$1 $\textrm{m}^3$ of sample volume at 10 mK temperature in a dilution refrigerator. The search will have sensitivity to axions and hidden photons over a wide range of mass and coupling\cite{PRD}. In this paper, we discuss the experimental design of the initial DM Radio Pathfinder experiment.

\section{Method of Detection}

While the full DM Radio experiment will probe both axions and hidden photons, the DM Radio Pathfinder is designed to probe only hidden photon dark matter, which does not require a large dc magnetic field for detection\cite{ADMX}. Hidden photons interact with photons via a kinetic mixing interaction. The strength of this interaction is parametrized by a mixing angle $\varepsilon$, which is known to be less than $10^{-6}$. We are interested in probing hidden photons of mass below 1 meV, for which the local number density is extremely high--over $10^{14}$ particles per cubic centimeter. Therefore, the hidden photon field may be represented as a classical vector field--in particular, a time-varying effective current density field permeating all space. The oscillation frequency is determined by the mass: $\nu_{\gamma '}= m_{\gamma '}c^{2}/h$, where $m_{\gamma '}$ is the hidden photon mass, $c$ is the speed of light, and $h$ is Planck's constant. Because of the kinetic energy of the hidden photon, which is set by the dark matter virial velocity $10^{-3} c$, the bandwidth of the hidden-photon signal is $10^{-6} \nu_{\gamma '}$. Furthermore, the virial velocity endows the current density field with a macroscopic coherence length $1000c/\nu_{\gamma '}$, over which the field is spatially uniform. For a 1 meV hidden photon, this coherence length is 1.2 meters and longer than any detector that we discuss in this paper. As such, we can treat the hidden-photon field as being uniform within our detector volume.

\begin{figure}
\centering
\includegraphics[width=3.5in]{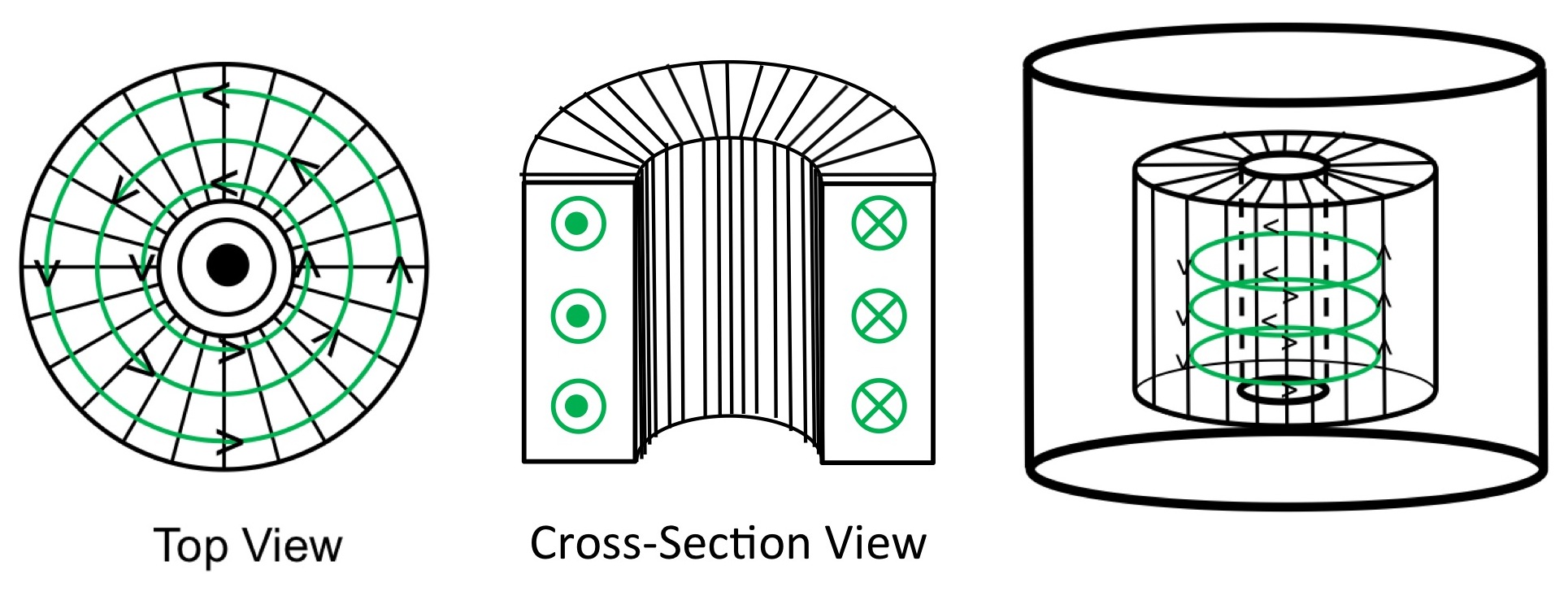}
\caption{ \small The magnetic field generated in the closed toroidal sheath is shown in top view (left), cross-sectional view (middle), and front view (right). The front view (right) also shows the superconducting shield that blocks external electromagnetic interference --- this shield is not shown in the left or middle view.}
\label{Fields in Res}
\end{figure}

\begin{figure}
\centering
\includegraphics[width=3.5in]{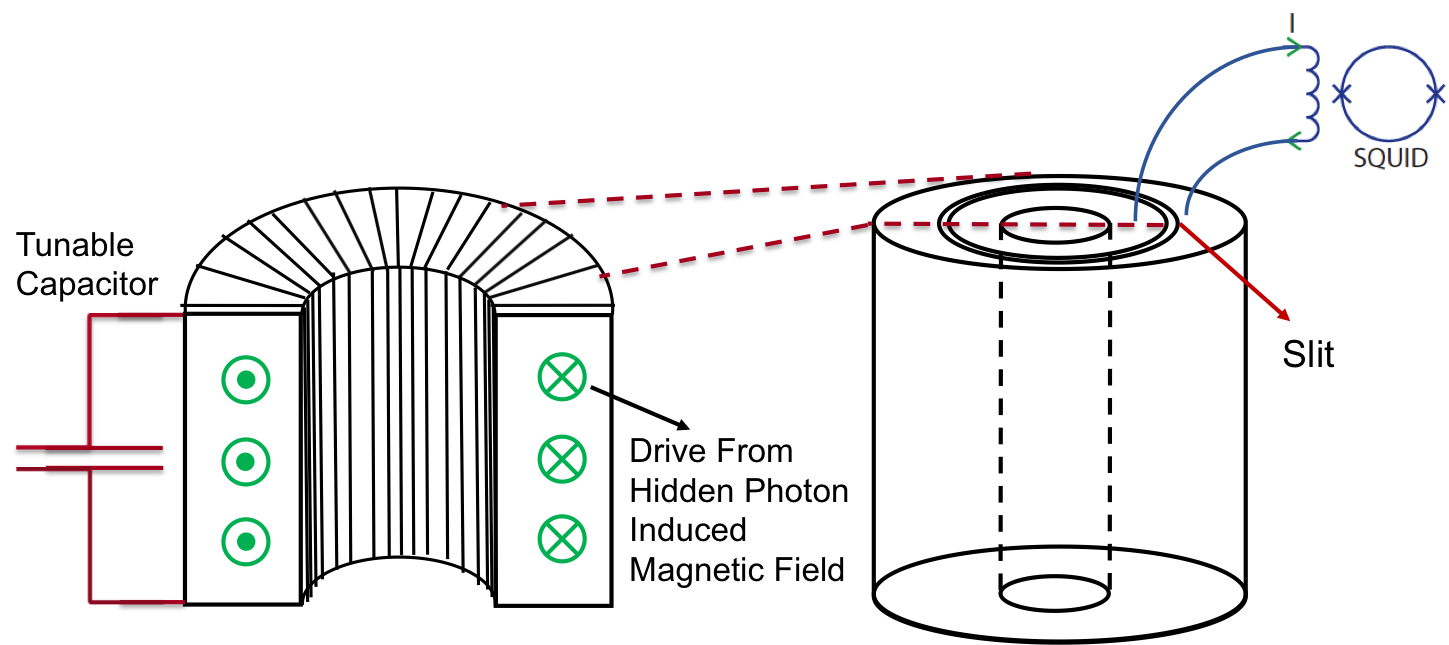}
\caption{ \small To increase the signal-to-noise ratio of the signal, a superconducting solenoidal inductor coil with (screened) inductance $L$ is wrapped around the toroidal pickup sheath, and connected in series to a superconducting tunable capacitor of value $C$ (left). The sheath has a slit cut circumferentially at the top; screening currents are inductively coupled to a SQUID that is connected across this slit (right). This high-$Q$ circuit has a resonance $\nu_0 = \frac{1}{2\pi\sqrt{LC}}$. If a hidden-photon field drives this circuit on resonance, the resonator rings up, increasing the amplitude of the screening current driven through the SQUID.}
\label{Resonant & SQUID circuit}
\end{figure}

A superconducting shield encloses the detector. The cylindrical shield blocks external electromagnetic fields, but is penetrated by the hidden photon field. The hidden photon effective current density produces a circumferential magnetic field inside the shield.
The magnetic field generated by the hidden photon effective current is shown in Fig. \ref{Fields in Res} in the case that the detector axis is aligned with the effective current. A toroidal sheath with a circular slit (shown at the top in Fig. \ref{Resonant & SQUID circuit}) couples to this field. A SQUID attached across the slit senses the screening currents. This system is, in principle, sensitive to hidden photons across a broad bandwidth; however, the signal to noise ratio is low. To increase the sensitivity, a solenoidal inductor is wrapped around the sheath, and connected in series to a parallel-plate capacitor. (Fig. \ref{Resonant & SQUID circuit}) When the resonance frequency equals the hidden photon frequency, the resonator rings up the screening current, increasing the signal to noise ratio. We target a quality factor of $Q \sim 10^{6}$, matching the resonator bandwidth with the dark matter signal bandwidth. A tunable LC resonator may be realized by changing the number of turns in the solenoid or by adjusting the position of an insertable sapphire dielectric between the capacitor electrodes. Below, we discuss in more detail the various components of the detector design: the resonant detector, the tuning procedure, the SQUID readout, the shielding, and the cryogenic mounting structure. For a detailed discussion of the general design constraints (not specific to the Pathfinder) and experiment sensitivity, see \cite{PRD}.

\section{Prototype Design}

\begin{figure}
\centering
\includegraphics[width=3.5in]{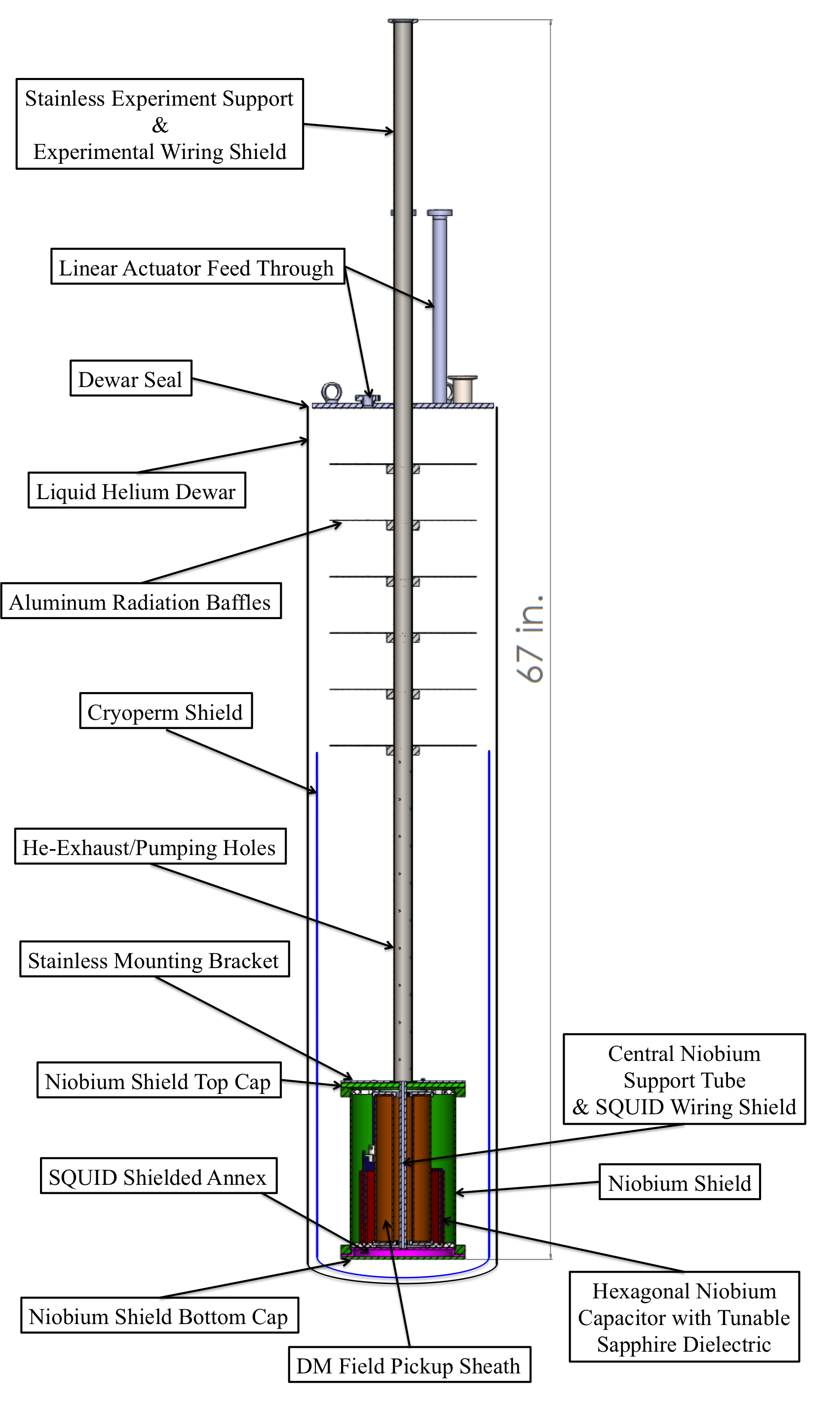}
\caption{\small A cross-section view of the DM Radio Pathfinder. The stainless-steel top portion supports the Nb shield in a liquid helium dewar.}
\label{Full Probe}
\end{figure}
The full prototype instrument is shown in figure \ref{Full Probe}. The detection circuit, consisting of a slitted pickup sheath and a tunable LC resonator, is located inside the superconducting shield at the bottom of the probe.

The shield is attached to a stainless steel mounting structure which is used to secure the experiment in a cryoperm-lined liquid helium dewar. This mounting structure and the shielding are described in section IIIB. The wiring and readout electronics have been designed to minimize readout noise and pickup. The cold readout circuitry, including a SQUID amplifier, will be housed in a separate shielded annex below the experiment (pink). The annex, SQUID, and readout chain are discussed in section IIIC.

\subsection{Tunable Resonator and Slitted Pickup Sheath}
The design of the resonator and slitted pickup sheath is driven by the need for high $Q \sim 10^{6}$ and the simultaneous need for precision tuning at the level of one part in $Q$ and large frequency throw (a factor of 3 in frequency between inductor coil set changes). 

\begin{figure}
\centering
\includegraphics[width=3.5in]{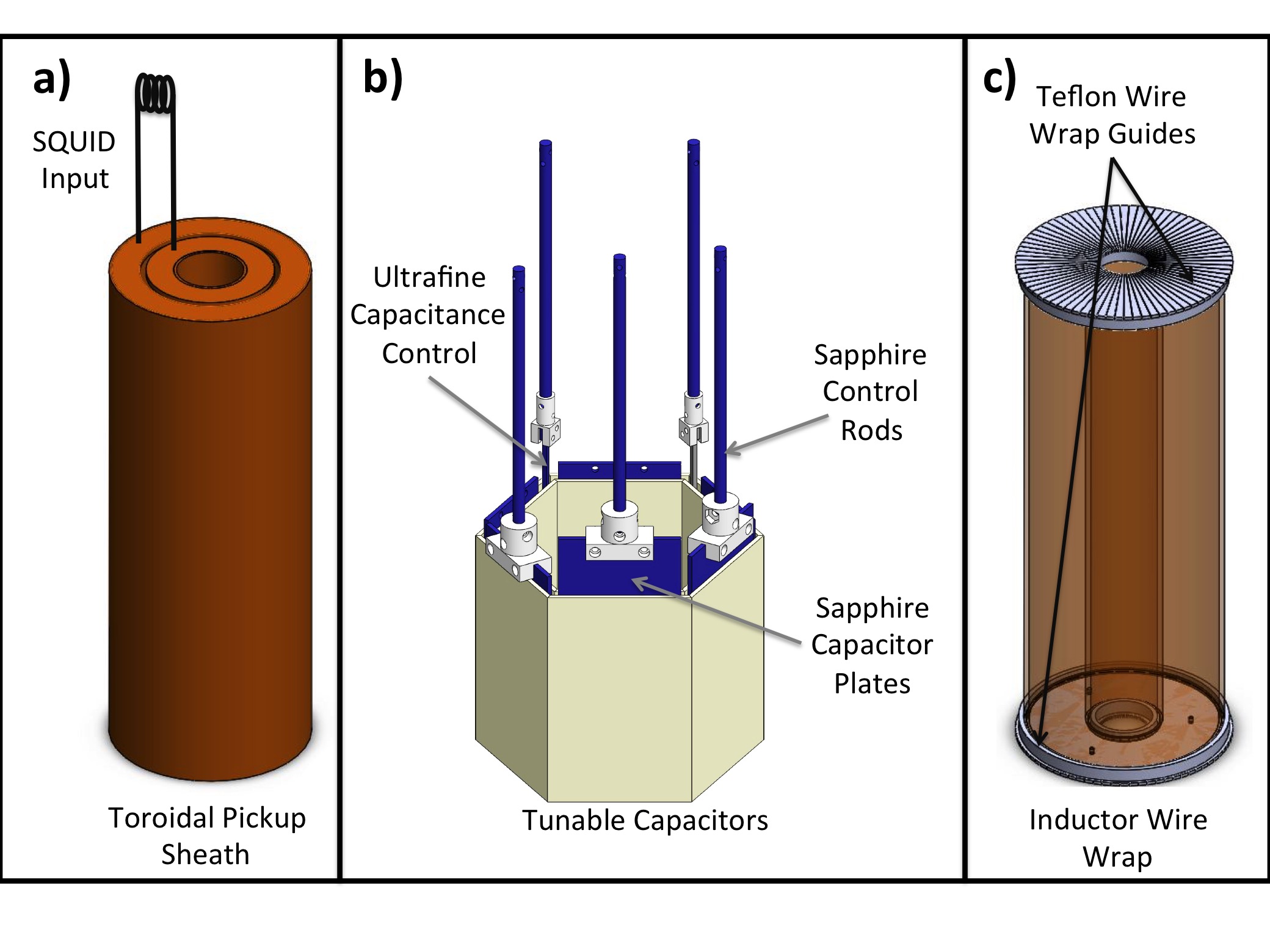}
\caption{ \small Components of the DM Radio Pathfinder. a) A slitted pickup sheath, which is made from e-beam welded niobium pieces. The SQUID input leads are spot welded across the circumferential slit in the top of the sheath. b) An array of six tunable capacitors. The rectangular capacitor plates are aligned in the shape of a hexagonal prism, and are electrically connected in parallel. The capacitors use sapphire as a dielectric material. They are tuned by sapphire control rods that connect to room- temperature linear actuators to adjust the amount of dielectric in the electric field between the plates, and thus the resonant frequency. c) A solenoidal inductor wrapped around the pickup sheath. The inductor is wired in series with the tunable capacitor. The toroidal solenoid is wrapped around small machined PTFE wire-wrap guides that fit over the ends of the pickup sheath.}
\label{Sub Components Resonator}
\end{figure}

In the Pathfinder experiment, the slitted pickup sheath (Fig. \ref{Sub Components Resonator}a) is realized as a hollow form made from 2mm thick, 99.9 \% pure Nb.  Niobium is chosen for its relatively high $T_{c}$ (9.2 K) and compatibility with high $Q$ at liquid helium temperature ($\approx$ 4.2 K at room pressure, $\sim$ 1 K if the helium bath is pumped). The sheath is 20.1 cm tall, with inner and outer diameter of 2.9 cm and 7.2 cm, respectively. This yields a sheath inductance of 36 nH. The wire leads that go to the input coil of the SQUID in the annex are spot welded across the slit. These wires are NbTi, which is also superconducting at liquid helium temperatures; we will experiment with both bare and formvar-coated wire. While the former increases the risk of shorts in the detector, the latter may degrade the $Q$ of the resonator via dielectric dissipation in the insulation. These tradeoffs must be understood in the course of the experiment.

The LC resonator is inductively coupled to the pickup sheath. The inductor $L$ is realized by winding a NbTi solenoid around the sheath. We will test coils with both bare and formvar-coated wire. To prevent shorts to the sheath and to facilitate wrapping the wire around the sheath, a PTFE wire guide is inserted between the sheath and the solenoid. This is shown in figure \ref{Sub Components Resonator}c. PTFE is chosen for its good dielectric loss properties and for ease of machining. The resonator capacitor $C$ is realized as a hexagonal parallel plate capacitor as shown in figure \ref{Sub Components Resonator}b. Plates of dimensions 5.6 cm $\times$ 9 cm  are made from 2 mm thick 99.9\% Nb (same as the sheath). These plates are then electron-beam welded to form the hexagonal shape. Electron-beam welding does not use lossy metal fillers that would degrade the $Q$. The separation of the hexagonal electrodes is 3 mm. To change the capacitance, we insert sapphire dielectric. The 99.9\% pure sapphire is oriented such that the C-axis is along the direction of the electric field in the capacitor. The dielectric constant along the C-axis is 11.5, which permits a frequency throw of $\sim$$\sqrt{11.5} \times \approx 3.4 \times$. Sapphire has an extremely low loss tangent ($<10^{-6}$), enabling high $Q$. The sheath and LC resonator are centered in the superconducting shield using PTFE centering brackets.

We use a capacitance tuning process with four levels of precision: ``ultra-fine'', ``fine'', ``coarse'', and ``ultra-coarse''. For the ultra-fine tuning, we insert a 2.5 cm Nb needle of 2 mm $\times$ 2 mm square cross-section between the hexagonal capacitor electrodes. A 25 $\mu$m change in the depth of the insertion of the needle between plates corresponds to a fractional change in capacitance of slightly less than one part in $10^6$. For the fine tuning, we adjust a 3.4 cm sapphire needle of 2 mm $\times$ 2 mm square cross section from the space between electrodes. A 25 $\mu$m change in the position of the needles corresponds to a fractional change in capacitance of approximately 1 part in 50,000. For the coarse tuning, we adjust a sapphire dielectric plate of dimensions 5.1 cm $\times$ 10.2 cm between electrodes. A 25 $\mu$m change in the depth of penetration of the dielectric between the plates corresponds to a fractional change in capacitance of approximately 1 part in 2500. For initial tests, the sapphire plate thickness is 2 mm, leaving a 500 $\mu$m separation between the dielectric and niobium plate, reducing alignment challenges. After the alignment procedures have been developed, thicker sapphire dielectrics will be substituted for a larger frequency adjustment range. There is one sapphire plate for each face of the capacitor electrode (six in total). Three of the six electrodes can be adjusted while cold.
For the ultra-coarse tuning, the other three sapphire plates are adjusted or removed between cooldowns. For the ultra-fine, fine, and coarse tuning, we control the position of the needles and plates using linear actuators located on the stainless steel top plate. The sapphire is coupled to the linear actuators by a set of rods.

The capacitive tuning allows a factor of 3 throw in frequency. To get a larger change in frequency, we change the number of turns in the toroidal solenoid inductor. By using as few as 8 turns and as many as 225 turns, we sweep the entire 100 kHz-10 MHz band (mass range 500 peV-50 neV) probed in the Pathfinder experiment.

\subsection{Shield and Mounting Structure}

For a high fidelity hidden photon search, we must shield the resonant detector from external fields that would otherwise present as electromagnetic interference. We use two layers of shielding: (1) a superconducting Nb shield to block AC fields and (2) a cryoperm shield lining the liquid helium dewar to reduce the DC field and flux trapping in the superconductors.

\begin{figure}
\centering
\includegraphics[width=3.5in]{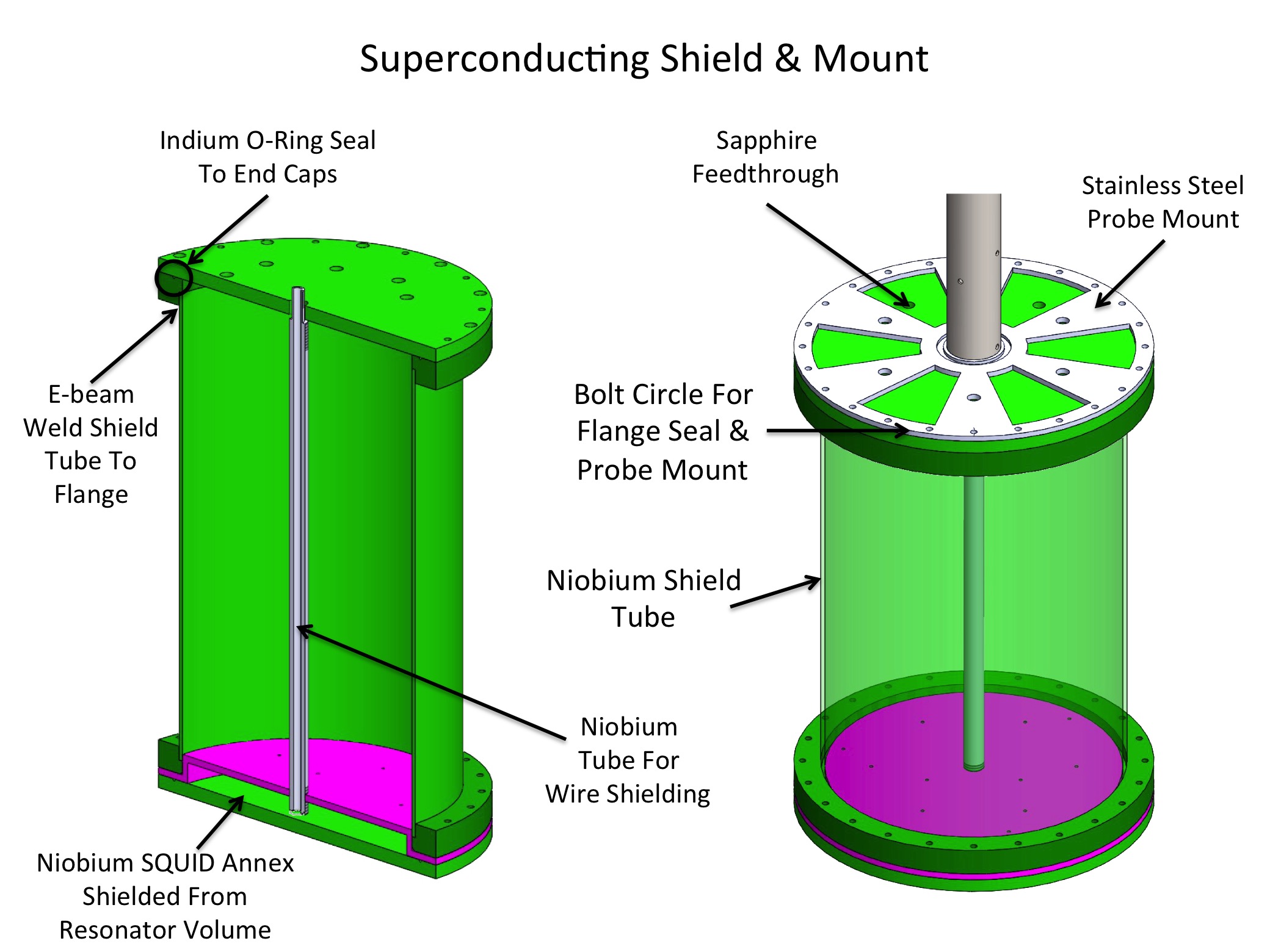}
\caption{ \small A superconducting RF shield surrounds the detector to block electromagnetic signals from exciting the LC oscillator. The shield is mounted to a stainless steel probe structure used to suspend the detector in a helium dewar. The shield is made from niobium and consists of a main cylinder/tube with flanges on both ends that have a bolt circle and indium o-ring groove to seal to top and bottom caps. In between the bottom flange and bottom cap is an annex where the cryogenic readout electronics are mounted to isolate them from the detector's shielded volume.}
\label{Shield}
\end{figure}

The cylindrical superconducting shield is shown in Fig. \ref{Shield}. The cylinder has a wall thickness of 2 mm and an outer diameter of 17 cm. Flanges are e-beam welded to the ends of the tube. Each flange contains a v-groove around its perimeter, in which we will seat an indium O-ring. This O-ring will be used to seal end caps to the cylinder. At the top of the shield, a bolt circle mechanically attaches the end cap and flange to the stainless steel mounting structure. A second bolt circle at the bottom attaches the flange to the SQUID annex discussed in Section IIIC and the bottom end cap. The bottom end cap contains holes for draining liquid helium from the shield. The holes are small enough to block electromagnetic signals at all frequencies of interest.	

A stainless-steel mounting structure is used to raise and lower the detector apparatus into the liquid helium bath. The top plate contains the linear actuators for tuning, ports for dewar wiring (thermometry, helium level sensing, and capacitive discharge wiring), ports for SQUID wiring, and ports for liquid helium transfer, venting, and pumping to reduce the bath temperature. A thin-walled (wall thickness 0.8 mm) stainless steel tube runs from the top plate. A stainless-steel plate at the bottom of the tube supports the shield.

\subsection{Readout}

The signal is amplified by a differential SQUID sensor to mitigate common-mode environmental pickup. \cite{DiffSQUID} Two 32-element SQUID series arrays from Magnicon GmbH are wired so that the SQUID output carries a differential signal. This is achieved by chaining the input coils so that the two series arrays receive signal flux of opposite polarity. The total input inductance of the SQUID is 12 nH, which is a reasonable match to the 36 nH sheath inductance. The white-noise level of the SQUID is below 3 pA/$\sqrt{Hz}$. Wires from the sheath are routed into a separate shielded annex (magenta in Fig. \ref{Shield}) containing the SQUID to prevent parasitic interactions with the resonant circuit. The SQUID output is carried on wires through a narrow Nb tube (center rod in graphic) and into the stainless steel tube. The signal is read out at room temperature with a low-noise, high gain differential amplifier made by Magnicon.

\section{Conclusion}
The DM Radio Pathfinder is now being assembled. With a total scan time of three months (and substantially more time in setup and change of coil sets), Pathfinder will probe hidden photon dark matter between 100 kHz and 10 MHz, down to mixing angles as low as $10^{-11}$. The techniques learned during the Pathfinder experiment will inform the construction of the full DM Radio experiment. 

The full DM Radio experiment will instrument $\sim$1  $\textrm{m}^3$ of sample volume at 10 mK in a dilution refrigerator. It will use dc SQUID amplifiers at the lowest frequency, dissipationless ac SQUIDs\cite{ACSQUID} at mid-frequencies, and parametric amplifiers at higher frequencies. The full DM Radio will incorporate a superconducting magnet, and will be sensitive to both axions and hidden photons. In addition to being more sensitive, DM Radio will cover a larger frequency range.

\section*{Acknowledgments}
We would like to thank Gary Sloan at Santa Clara University for
machining the stainless steel probe structure and the Stanford Physics Department
machine shop staff for niobium machining. The design of the DM Radio Pathfinder experiment was supported by the Kavli Institute for Particle Astrophysics and Cosmology (KIPAC), and the construction and operation of the DM-Radio Pathfinder Experiment is supported by the Department of Energy, Laboratory Directed Research and Development funding. Arran
Phipps acknowledges support from KIPAC through a Kavli fellowship.

\end{document}